\documentclass{article}
\usepackage[utf8]{inputenc}
\usepackage{url}

\title{Crecimiento con alta desigualdad implica p\'erdida de bienestar: Un modelo matem\'atico\footnote{A los inquietos estudiantes. Art\'iculo en construcci\'on.}}
\author{Fernando Córdova-Lepe}
\date{November 2019}

\usepackage{graphicx}

\begin{document}

\maketitle

\begin{abstract}
Un modelo matem\'atico de medida de la percepci\'on propia del bienestar para grupos con ingresos crecientes, pero proporcionalmente desiguales es propuesto. Asumiendo que el bienestar crece con el ingreso propio y decrece con la desgualdad relativa (ingreso del otro respecto al propio), se concluyen posibles escenarios para el comportamiento de largo plazo en las funciones de bienestar. Adem\'as, se prueba que una desigualdad relativa alta (definici\'on param\'etrica), siempre implica la p\'erdida de la auto percepci\'on del bienestar del grupo m\'as desfavorecido.
\end{abstract}
\vspace{5mm}

{\it La igualdad no significa que todos tengamos la 
misma riqueza, sino que nadie sea tan rico como para poder comprar a otro y nadie sea tan pobre como tener que venderse.} {\bf Jean-Jacques Rousseau}

\section{Introducci\'on}

Chile, desde mediados del mes de octubre del presente a\~no, vive un estallido masivo de profundo descontento social. Al respecto, uno de los primeros puntos que mencionan las agencias de noticias para comenzar sus relatos y an\'alisis, ha sido la paradoja de presenciar la revuelta esperable en sociedades en recesi\'on o con bajo crecimiento, tambi\'en en un pa\'is en que las cifras macroecon\'omicas muestran ingresos crecientes y superaci\'on de cierta l\'inea de la pobreza de buena parte de la poblaci\'on. {\it Violento fin de semana en Chile desnuda las desigualdades en la ``Suiza" de Latinoamérica}, 21/10/2019, Reuters\cite{R}. 
{\it Chilenos anticipan más protestas, hartos de la desigualdad}, 28/10/2019, Associated Press\cite{AP}. {\it `Chile despert\'o': el legado de la desigualdad desata protestas masivas}, 4/11/2019, The New York Times\cite{NYT}.
\vspace{2mm}

Rolf Luders, el caso cero (\'indice) en la propagaci\'on del ideal ``neoliberal" de M. Friedman en Chile, en una entrevista del 06/11/2019 a BBC News-Mundo\cite{BBC}, explicando la impopularidad al actual modelo dice: ``{\it ..., a pesar del enorme progreso, los ingresos de la gran mayoría de los chilenos aún son bajos en términos absolutos y las diferencias de ingreso son muy significativas}". Adem\'as, a\~nade a la defensa de la estrategia econ\'omica el siguiente reconocimiento: ``{\it No obstante, al privilegiar absolutamente el crecimiento y así a la reducción de la pobreza sobre la disminución de la desigualdad, se malinterpretaron las preferencias ciudadanas}".
\vspace{2mm}

Chile est\'a dentro de las econom\'ias que, tras un sostenido aumento en el ingreso de todos los sectores a la vista sin \'exito en cuanto a generar paz social, han mantenido en contraste una muy desigual distribuci\'on de la riqueza. Estamos evidenciando que crecimiento sin igualdad no es sostenible. ``{\it Las protestas estallaron por el incremento del boleto de Subte. Pero en realidad, y a pesar del buen nivel económico chileno, hay un malestar más profundo por la desigualdad y el costo de los servicios y la salud para los sectores más pobres}", 19/11/2019, Clarin Mundo\cite{CLARIN}. Advertencias respecto al peligro econ\'omico que conlleva la desigualdad, como un elemento capaz de afectar el crecimiento ya exist\'ian, ver \cite{RGP}. En \cite{JEN} leemos: ``{\it ... es generalizado el reconocimiento de que la desigualdad disminuye el crecimiento y la desigualdad creciente lo afecta de manera multiplicada}". Hacia mediados de 2004, Mario Kremerman, Fundaci\'on Terram\cite{MK}, titula un art\'iculo: {\it Distribuci\'on del Ingreso en Chile: Una Bomba de tiempo}.
\vspace{2mm}

Un estudio de
Michael Förster y Celine Thevenot\cite{LM}, economistas del Centro para la Oportunidad y la Igualdad de la OCDE, demostrar\'ia que: ``{\it la desigualdad de ingresos genera un fuerte impacto sobre el crecimiento económico y sobre el desarrollo social en general. Efectivamente, la desigualdad en la distribución de la riqueza en un país tiene efectos económicos, sociales, éticos y políticos negativos, al punto de restringir el crecimiento y afectar la confianza en las instituciones}". Es lo que, para el criterio de muchos especialista, habría estado ocurriendo en Chile, pero sin mayor reacci\'on de parte del poder pol\'itico y económico. 
\vspace{2mm}

Entendiendo la percepci\'on de la desigualdad en la distribuci\'on social del ingreso como uno de los elementos importantes en la variaci\'on en el tiempo del bienestar propio, el presente trabajo apunta a dar una explicaci\'on deductiva simple respecto a la p\'erdida de bienestar de las personas (las desfavorecidas por la desigualdad) aun bajo la hip\'otesis de un crecimiento en el ingreso. El lenguaje y la metodolog\'ia, para la derivaci\'on de conclusiones, tiene la perspectiva  de la modelizaci\'on matem\'atica. M\'as precisamente,
instalamos matematicamente dos afimaciones: (a) las variaciones en el ingreso correlacionan positivamente con variaciones en el bienestar y, adem\'as, (b) la comparaci\'on relativa del ingreso de los otros respecto al propio (criterio de comparaci\'on) es un elemento de p\'erdida de la auto-percepci\'on del bienestar. 
\vspace{2mm}

El concepto de bienestar de una persona o de un segmento de la sociedad es claramente subjetivo, aunque el ingreso es un componente importante, \'este por s\'i solo no es del todo determinante. En \cite{N-S}, se reconoce que {\it las comparaciones sociales en ingreso
y los contactos sociales pueden ser determinantes claves del bienestar subjetivo}. Hay literatura que muestra que el bienestar de un individuo no solo depende de su propia situaci\'on, sino de lo que observa a su alrededor. En este sentido, puede depender de:
(i) el propio ingreso actual, (ii) el ingreso en el pasado y (iii) de la comparación con los ingresos de otras personas, ver \cite{A-F,F-C,OIDLES}. El modelo que proponemos considera m\'as directamente las posibilidades (i) y (iii). 
\vspace{2mm}

El bienestar personal tiene una correlaci\'on directa con su ingreso solamente si se hace el ejercicio mental (artificial) de aislar el resto de los m\'ultiples factores involucrados, en particular respecto a la variable desigualdad. Est\'a probado estad\'isticamente, en estudios de corte tranversal, ver \cite{E}, que ``{\it ... la relación entre ingreso y felicidad es positiva y estad\'isticamente significativa, pero d\'ebil en el sentido de que aumentos considerables en el ingreso tienen, en promedio, un efecto relativamente peque\~no en la felicidad".}
El mismo estudio se\~nala que hay una probabilidad no despreciable que existan personas de ``{\it ingreso alto con baja felicidad y personas de bajo ingreso con alta felicidad no es despreciable}".
\vspace{2mm}

Nuestro objetivo es obtener elementos de juicio, desde la perspectiva que puede ofrecer un modelo matem\'atico simple, pero que incorpora al menos las variables ingreso, desigualdad y bienestar. Esto es, condiciones entre los pa\-r\'a\-me\-tros del modelo que determinan crecimiento y/o decrecimiento del bienestar de dos grupos con ingresos desiguales pero crecientes.

\section{El Modelo}

La tercera acepci\'on que ofrece la Real Academia Española para la palabra {\it bienestar} es ``{\it Estado de la persona en el que se le hace sensible el buen funcionamiento de su actividad somática y psíquica}". Es claro que se refiere a una definici\'on en positivo de bienstar personal, que asume cubiertas y en equilibrio tanto el aspecto f\'isico como el emocional del individuo, lo que implica un alto grado  de subjetividad. Notemos que al vivir en sociedad, nuestro equilibrio emocional depende fuertemente de lo que ocurre con los otros que est\'an en nuestro entorno, una vecindad física o del espacio informativo.
\vspace{2mm}

Notemos que en el esfuerzo de substraerse de lo individual, las definiciones m\'as cl\'asicas del concepto de {\it bienstar social} se concentran m\'as en el aspecto material, en el sentido que aparecen relacionados con algunos factores econ\'omicos objetivos. En este trabajo, dejamos de lado la idea de un individuo socialmente uniforme y se incorpora (y aqu\'i puede residir la mayor novedad) el supuesto de una sociedad segmentada por ingresos en la que un indicador del bienestar del segmento depende del ingreso propio y tambi\'en, por comparaci\'on, del ingreso de los otros. Adem\'as, entendiendo por bienestar, no un estado positivo {\it a priori}, sino como una variable medible.
\vspace{2mm}

Asumiremos, por simplicidad, que una comunidad de intereses ({\it v.g.}, un pa\'is) est\'a segmentada por nivel de ingresos en dos grupos, que denotaremos por $G$ y $G^{*}$. Adem\'as, consideramos que la autopercepci\'on  del bienestar de un grupo es representable por una variable num\'erica de rango continuo no negativa. Variable que al cambiar a un valor num\'erico mayor, es signo de m\'as bienestar. Denotaremos estas variables de bienestar para $G$ y $G^{*}$,
como funci\'on del tiempo, por $B(\cdot)$ y $B^{*}(\cdot)$ respectivamente.
\vspace{2mm}

Otra hip\'otesis es suponer
que las variables que definen los determinantes principales para la percepci\'on del bienestar de cada uno de los grupos son los ingresos monetarios (corregidos por inflaci\'on) del grupo en cuesti\'on y tambi\'en del grupo complementario.
Representaresmos los ingresos individuales, por unidad de tiempo, como variables no negativas de rango continuo en el tiempo, las que denotamos por funciones $p(\cdot)$ y $q(\cdot)$, respectivamente para las personas de $G$ y $G^{*}$.
\vspace{2mm}

El vivir en sociedad implica el tener a disposici\'on, como informaci\'on de todos los actores, alguna aproximaci\'on de la funci\'on de ingreso del resto o de los subgrupos sociales m\'as visibles, sea en t\'erminos objetivos o seg\'un los intereses, donde enfoca la atenci\'on, el grupo al que nos referimos. As\'i, la percepci\'on que tiene el grupo referente $G$ respecto del ingreso del grupo complementario $G^{*}$ y viceversa (normalizada por tama\~no), se asumirá absolutamente conocida.
\vspace{2mm}

As\'i, obviando una serie de din\'amicas sociales ({\it v.g.}, demogr\'aficas, de flujo entre los grupos y de cambios conductuales en el consumo), vamos a considerar que la autopercepci\'on del bienestar en cada segmento $G$ y $G^{*}$, var\'ia principalmente por dos factores o criterios: 
[f$_{1}$] {\it cambios en
 ingreso propio}, como elemento que suma bienestar y [f$_{2}$] {\it comparaci\'on respecto al ingreso del grupo complementario}, como componente que substrae. Esto es, dado un instante $t$ y una variaci\'on temporal $\Delta t>0$, tenemos que $\Delta B=B(t+\Delta t)-B(t)$, est\'a determinado por:
$$
\label{B0}
 \Delta B/B = 
 \mathcal{F}_{1}\left[\begin{array}{c} 
                  \mbox{Cambio en ingreso propio}\\
                  \mbox{en}\,\,[0,\Delta t]\\
                  \end{array} \right]
-\mathcal{F}_{2}\left[\begin{array}{c}
                  \mbox{Comparaci\'on con ingreso}\\
                  \mbox{otros en}\,\, [0,\Delta t]\\
                  \end{array} \right].   
$$
Esto es, el cambio relativo (o porcentual) de la percepci\'on del bienestar propio 
$\Delta B/B$
sube con $\mathcal{F}_{1}[\Delta p]$ y disminuye con $\mathcal{F}_{2}[p,q]$. Lo que se asume igualmente para el grupo $G^{*}$, pero en t\'erminos sim\'etricos, para $B^{*}(\cdot)$. \vspace{2mm}

En orden a generar conclusiones a bajo costo t\'ecnico, introduciremos una especificaci\'on funcional simple de estos factores:

\begin{itemize}
    \item[f$_{1}$] {\it Cambio en ingreso propio}: Una variaci\'on porcentual en el ingreso implica una variaci\'on porcentual proporcional en la percepci\'on de las funciones de bienestar. Esto es, un cambio monetario $\Delta p$ en el ingreso del grupo referido, determina un aumento en una cantidad
    $ a\, \Delta p/ p$ en la autopercepci\'on del bienestar. Esto es,
    $$
    \mathcal{F}_{1}\left[\begin{array}{c} 
                  \mbox{Cambio en ingreso propio}\\
                  \mbox{en}\,\,[0,\Delta t]\\
                  \end{array} \right]=\mathcal{F}_{1}[\Delta p]=a\,\Delta p/p,
    $$
    donde $a$ es una constante positiva de proporcionalidad, una sensibilidad, que mide el aumento relativo en bienestar ante una variaci\'on del ingreso en una unidad monetaria. Notemos que este supuesto contrasta con el cl\'asico trabajo de Dalton\cite{Dalton} que al asumir {\it la hipótesis de que las adiciones proporcionales a los ingresos, superiores a las requeridas para una ``subsistencia simple", hacen adiciones iguales al bienestar económico}, en nuestros notaci\'on se escribir\'ia $\Delta B = \Delta p /p $.
    \vspace{2mm}

    \item[f$_{2}$] {\it Comparaci\'on con ingreso otros}: Para  el grupo objeto existe una tasa $b$ de p\'erdida de bienestar por unidad de tiempo (desgaste natural) cuando los ingresos $p$ y $q$ son iguales. Sin embargo, en t\'erminos comparativos, esta proporci\'on ser\'a mayor si el ingreso del otro es mayor y es menor si la del otro es menor. Asumiremos as\'i una p\'erdida porcentual en el bienestar $B$, por unidad de tiempo, en una fracci\'on $b\,q/p$. Esto es,
    $$
    \mathcal{F}_{2}\left[\begin{array}{c}
                  \mbox{Comparaci\'on con ingreso otros}\\
                  \mbox{en}\,\, [0,\Delta t]\\
                  \end{array} \right]=\mathcal{F}_{2}[p,q]=b\,\Delta t \, (q/p). 
    $$
    De modo que, si $p \ll q$ (mucho menor) o $p \gg q$ (mucho mayor) este sustraendo del indicador del bienestar es respectivamente considerablemente grande o peque\~no.
    \vspace{2mm}
    
    En relaci\'on al cuociente $q/p$, observemos que su valor relativo $[[q/p]]$ (introducido en \cite{FCL}, donde $[[x]]=x$, si $x \geq 1$ y $[[x]]=x^{-1}$, si $0<x<1$) satisface el {\it Principio de transferencias} (Axioma 1), la {\it Simetr\'ia de poblaci\'on} (Axioma 2) y  la {\it Invarianza de escala} (Axioma 3), propios para una teor\'ia b\'asica de la medida de la desigualdad, seg\'un \cite{JS-CW}.
    \end{itemize}

Sintetizando, nuestras hip\'otesis simples (f$_{1}$ y f$_{2}$) de regulaci\'on del cambio en el bienestar auto\-per\-ci\-bido, est\'a dada por: 
\begin{equation}
\label{B}
\Delta B =\left\{\mathcal{F}_{1}[\Delta p] -\mathcal{F}_{2}[p,q]\right\}\,B=\left\{a \frac{\Delta p}{p}- b\,\Delta t\,\frac{q}{p} \right\}\,B.
\end{equation}

Para explicitar el tiempo y facilicitar c\'alculos, podemos escribir (\ref{B}) en t\'erminos diferenciales  dividiendo, por $\Delta t$,
y tomando el l\'imite cuando $\Delta t$ tiende a cero. Obte\-ne\-mos:
\begin{equation}
\label{B2}
\frac{dB}{dt}(t) =\left\{a\,\frac{p'(t)}{p(t)}-b\, \frac{q(t)}{p(t)}\right\}\,B(t).
\end{equation}

Notemos que la ecuaci\'on (\ref{B2}) de la autopercepci\'on del grupo $G$ tiene su correlato para el grupo $G^{*}$ con respectivos par\'ametros $a^{*}$, $b^{*}$ e intercambiando las funciones de ingreso $p(\cdot)$ y $q(\cdot)$. 
Esto es,
\begin{equation}
\label{B2*}
\frac{dB^{*}}{dt}(t) =\left\{a^{*}\,\frac{q'(t)}{q(t)}-b^{*}\, \frac{p(t)}{q(t)}\right\}\,B^{*}(t).
\end{equation}
\vspace{3mm}

Observemos que no es sostenible para los factores $a$, $a^{*}$, $b$ y $b^{*}$ que estos sean constantes en el tiempo, pero s\'i que en el promedio de los individuos por grupo, se mantengan en ciertos m\'argenes. En nuestro caso, se asumir\'an fijos, lo que es suficiente para obtener unas primeras conclusiones en dichos m\'argenes. 
\vspace{2mm}

Los factores $a$ y $a^{*}$ traducen cambios proporcionales en el ingreso individual en $G$ y $G^{*}$ respectivamente, en cambios proporcionales en su bienestar, asumiendo las otras variables fijas. Entonces,
si el margen de ingresos, por ejemplo para $G$, est\'a muy cercana o por debajo a alguna l\'inea de subsistencia se espera que $a$ sea mayor que cuando existe alguna holgura. Tampoco es ajeno a la cultura de una sociedad y las consideraciones de orden psicol\'ogico en la sensaci\'on de bienestar. Por ejemplo, si la medida predominante del \'exito personal est\'a indexado por la tenencia de bienes materiales, muy probablemente significa un factor $a$ mayor.
\vspace{2mm}

En cuanto a los factores $b$ y $b^{*}$, estos est\'an representando la p\'erdida por desgaste na\-tu\-ral del bienestar y se han supuesto se incrementan o disminuyen seg\'un la comparaci\'on de ingresos del otro respecto al propio. En este sentido, pueden ser mayor o menor seg\'un una serie de consideraciones econ\'omicas, sociales o psicol\'ogicas, como la influencia de ejes como patrimonio-escasez, resentimiento-conformidad o empat\'ia-indiferencia.

\section{An\'alisis del modelo}

Por simple integraci\'on en (\ref{B2}) y (\ref{B2*}) sobre un intervalo temporal $[t_{0},t]$ tenemos el despeje que sigue:
\begin{equation}
\label{B3}
B(t)=B_{0} \left[\frac{p(t)}{p_{0}}\right]^{a} e^{-b\int_{t_{0}}^{t}q(s)/p(s)\,ds
}\quad \mbox{y} \quad
B^{*}(t)=B_{0}^{*} \left[\frac{q(t)}{q_{0}}\right]^{a^{*}} e^{-b^{*}\int_{t_{0}}^{t}p(s)/q(s)\,ds
},
\end{equation}
para $t \geq t_{0}$, donde $(B_{0},p_{0})$ y 
$(B_{0}^{*},q_{0})$ son par\'ametros que denotan el par ({\it bie\-nes\-tar}, {\it ingreso}) inicial, es decir, $(B(t_{0}),p(t_{0}))$ y $(B^{*}(t_{0}),q(t_{0}))$ respectivamente a los grupos $G$ y $G^{*}$.
\vspace{2mm}

Vamos a suponer un escenario social en el que la funci\'on de ingreso $p(\cdot)$ del grupo objetivo $G$ es de tipo exponencial a una tasa positiva $\lambda$, mientras que  el ingreso del grupo complementario (en adelante $G^{*}$) satisface
$q(\cdot)=n\,p(\cdot)$, con $p(t)=p_{0}e^{\lambda(t-t_{0})}$, para cierto $n\geq 1$. Esto es, el grupo $G^{*}$ es el favorecido, $n$-veces m\'as ingreso que $G$, el grupo desfavorecido.  Denominaremos {\it desigualdad relativa de ingreso} al factor $n=[[q/p]]=[[p/q]]$.
\vspace{2mm}

As\'i, remplazando las funciones ingreso $p(\cdot)$ y $q(\cdot)$ en (\ref{B3}), la autopercepci\'on del bienestar  de los grupos $G$ y $G^{*}$, desde el momento $t_{0}$ en que el ingreso de $G$ es la $n$-\'esima parte de la del grupo $G^{*}$, quedan respectivamente definidas por:
\begin{equation}
B(t)=B_{0}\,\exp\left\{(a\lambda-b\,n)(t-t_{0})\right\} \quad \mbox{y}\quad 
B^{*}(t)=B_{0}^{*}\,\exp\left\{\left(a^{*}\lambda-b^{*}/n \right)(t-t_{0})\right\},
\end{equation}
para todo $ t \geq 0$.
\vspace{2mm}

Esta funciones exponenciales son mon\'otonas para $t\geq t_{0}$, crecientes, constantes o decrecientes, seg\'un el signo de $a\lambda-bn$ y $a^{*}\lambda-b/n$ sea positivo, cero o negativo. De modo que,
el comportamiento de largo plazo ({\it i.e.}, cuando $t\to \infty$) de las funciones $B(\cdot)$ y $B^{*}(\cdot)$ se puede resumir seg\'un el tama\~no de la tasa $\lambda$, del modo siguiente:

\begin{itemize}
    \item[(I)] Caso $\lambda^{2}<bb^{*}/(aa^{*})$:
    \begin{equation}
    \begin{array}{|l|c|c|c|c|}
    \hline
     & & & & \\
    \mbox{Desigualdad} & n & \,\, ]0,\frac{a\lambda}{b}[\,\, & \,\,]\frac{a\lambda}{b}, \frac{b^{*}}{a^{*}\lambda}[\,\,   & \,\,]\frac{b^{*}}{a^{*}\lambda},\infty[\,\,  \\
    & &\mbox{\footnotesize{baja}} & \mbox{\footnotesize{media}} & \mbox{\footnotesize{alta}} \\
    \hline
    \hline
    \mbox{Bienestar en} \quad G & B & \nearrow \infty & \searrow 0   &  \searrow 0\\ 
    & & & & \\
    \mbox{Bienestar en} \quad G^{*} & B^{*} & \searrow 0 & \searrow 0 & \nearrow \infty\\
     \hline
    \end{array}
    \end{equation}
    
     \item[(II)] Caso $\lambda^{2} > bb^{*}/(aa^{*})$:
    \begin{equation}
    \begin{array}{|l|c|c|c|c|}
    \hline
     & & & & \\
    \mbox{Desigualdad} & n & \,\, ]0,\frac{b^{*}}{a^{*}\lambda}[\,\, & \,\,]\frac{b^{*}}{a^{*}\lambda}, \frac{a\lambda}{b}[\,\,   & \,\,]\frac{a\lambda}{b},\infty[\,\,  \\
    & &\mbox{\footnotesize{baja}} &\mbox{\footnotesize{media}} &\mbox{\footnotesize{alta}} \\
    \hline
    \hline
    \mbox{Bienestar en} \quad G & B & \nearrow \infty & \nearrow \infty   &  \searrow 0\\ 
    & & & & \\
    \mbox{Bienestar en} \quad G^{*} & B^{*} & \searrow 0& \nearrow \infty & \nearrow \infty\\
     \hline
    \end{array}
    \end{equation}
    
\end{itemize}

Ahora, en orden a comparar $B(\cdot)$ con $B^{*}(\cdot)$, notemos que
\begin{equation}
\label{Q}
B(t)/B^{*}(t)=e^{-f(n)(t-t_{0})}B_
{0}/B^{*}_{0}, \quad t \geq t_{0},    
\end{equation}
con $f(n)=-[(a-a^{*})\lambda-(bn-b^{*}/n)]=bn^{2}-(a-a^{*})n-b^{*}$, un polinomio cuadr\'atico, que tiene discriminante positivo $\mathcal{D}=(a-a^{*})^{2}\lambda^{2}+4bb^{*}$, por lo que es factorizabe como
\begin{equation}
\label{f}
f(n)=b\left(n-\hat{n}\right)
\left(n+\{\sqrt{\mathcal{D}}-(a-a^{*})\lambda\}/\{2b\}\right),
\end{equation}
donde $
\hat{n}=\{(a-a^{*})\lambda+\sqrt{\mathcal{D}}\}/\{2b\}$.
\vspace{2mm}

Consideraremos que en (\ref{Q}), incialmente el nivel de bienestar es el mismo, i.e., $B_{0}=B^{*}_{0}$. En orden a comparar el bienestar de un grupo respecto a otro, notemos que el cuociente $B(\cdot)/B^{*}(\cdot)$ es menor, igual o mayor que uno seg\'un el signo de $f(n)$ 
sea positivo, cero o negativo. Es decir, ya que en (\ref{f})  tenemos $ \sqrt{\mathcal{D}}>(a-a^{*})\lambda$,  de acuerdo a si $n$ es menor, igual o mayor que $\hat{n}$.
\\

Entonces, de acuerdo a la tasa de crecimiento del ingreso, tenemos (prueba en el ap\'endice):
\begin{itemize}
    \item[(III)] Si $\lambda^{2}<bb^{*}/(aa^{*})$, entonces
    $\hat{n}\in\, ]a\lambda/b,\, b^{*}/(a^{*}\lambda)[$. As\'i, seg\'un la tabla en (I), en el nivel de desigualdad bajo $B(\nearrow)>B^{*}(\searrow)$. A desigualdad media y menor a $\hat{n}$, tenemos $B>B^{*}$, pero $B,\,B^{*}\searrow 0$. Ahora, a desigualdad media, pero mayor a $\hat{n}$, el caso es $B<B^{*}$ con $B,\,B^{*}\searrow 0$. Finalmente, a desigualdad alta $B(\searrow)<B^{*}(\nearrow)$.
    \vspace{2mm}
    
      \item[(IV)] Si $\lambda^{2}>bb^{*}/(aa^{*})$, entonces
    $\hat{n}\in \,] b^{*}/(a^{*}\lambda), a\lambda/b[$.
    Observando la tabla en (II), se tiene que
    con baja desigualdad, $B(\nearrow)>B^{*}(\searrow)$. A desigualdad media y menor a $\hat{n}$, tenemos $B>B^{*}$, pero ambas crecientes $B,\,B^{*}\searrow \infty$. Si la desigualdad relativa es media, pero mayor a $\hat{n}$, el caso es $B<B^{*}$ con $B,\,B^{*}\nearrow \infty$. A desigualdad alta $B(\searrow)<B^{*}(\nearrow)$.

\end{itemize}

\vspace{1cm}
\noindent{\bf Observaciones}:
\begin{itemize}
    \item[(i)] {\it Desigualdad relativa alta}: No importando el tama\~no de la tasa de crecimiento $\lambda$, es decir, en cualquiera de los casos I y II, si $n$ es suficientemente grande, entonces necesariamente la autopercepci\'on del bienestar el grupo $G$ decae a cero, mientras que el de $G^{*}$ crece sin cota.\\

    \item[(ii)] {\it Desigualdad relativa baja}: En cualquiera de los casos I y II, el bienestar de $G$ crece sin cota, pero la del grupo favorecido $G^{*}$ decae exponencialmente. Sin embargo, este caso es imposible cuando las tasas (sensibilidades) de los criterios f$_{1}$ y f$_{2}$ para los grupos $G$ y $G^{*}$ coinciden, esto es, si $a=a^{*}$ y $b=b^{*}$. En efecto, se requerir\'ia
$1 \leq n < \min\{a\lambda/b, (a\lambda/b)^{-1}\}<1$, lo que es una contradicci\'on.\\
    
    Notemos que tanto la condici\'on (I) $\lambda^{2}<bb^{*}/(aa^{*})$ junto a $1 \leq n <a\lambda/b$, como el caso II $\lambda^{2}>bb^{*}/(aa^{*})$ junto a $1 \leq n <b^{*}/(a^{*}\lambda)$, implican que se satisfaga la relaci\'on:
    $$
\frac{a^{*}\lambda}{b^{*}}<1<\frac{a\lambda}{b}.
$$ 
Desigualdad que se puede lograr asumiendo
$a^{*}\ll a\,\,$ y/o $\,\,b^{*}\gg b$. Es decir, el grupo favorecido $G^{*}$ debe ser con respecto a $G$, mucho menos sensible a los cambios en su ingreso y/o mucho m\'as sensible a la comparaci\'on del ingreso de $G$ respecto al suyo. \\
 
    \item[(iii)] {\it Desigualdad relativa media}: En el Caso I, que llamaremos de tasa de crecimiento baja, un $ n \in ]a\lambda/b, b^{*}/(a^{*}\lambda)[$, implica decaimiento del bienestar para ambos grupos. 
    Destaquemos que el \'unico caso en que ambos grupos pueden presentar una autopercepci\'on del crecimiento creciente es en el Caso II, de una tasa de crecimiento alta, esto es, por sobre e umbral $\sqrt{bb^{*}/(aa^{*})}$, supeditado a $n \in \mathcal{J}$, con 
    \begin{equation}
    \mathcal{J}=]b^{*}/(a^{*}\lambda),\,a\lambda/b[.    
    \end{equation}
    
    Observemos que $\mathcal{J}$, que denominaremos intervalo de {\it doble autopercepci\'on po\-si\-ti\-va del bienestar}, se ensancha ({\it resp.} angosta) con una tasa de crecimiento $\lambda$ mayor ({\it resp.} menor) del ingreso. Tambi\'en tenemos m\'as margen a la derecha si aumenta la sensibilidad a las variaciones del ingreso $a$ y la sensibilidad a la p\'erdida por comparaci\'on $b$. Inversamente, si disminuye $a^{*}$ y aumenta $b^{*}$ el margen izquierdo de $\mathcal{J}$ retrocede. Lo que nos da amplio espacio paraa posibilidades interpretativas. 
\end{itemize}

\section{Conclusiones}
A grandes rasgos, la 
hip\'otesis en el modelo, de un crecimiento exponencial en el ingreso ($\lambda>0$), se satisface en Chile. El PBI perc\'apita de las \'ultimas d\'ecadas muestra un crecimiento sostenido.
Seg\'un cifras del Banco Mundial,
este ha pasado de US\$ $5.064$ en el a\~no 2000 hasta US\$ $18.592$ para el 2018, con una proyecci\'on al 2019 de US\$ $21.190$. Adem\'as, en cuanto a su variaci\'on promedio por d\'ecada, las cifras del Banco Central de Chile, son: 3,6\% [1980-1989], 6,1\% [1990-1999], 4,2\% [2000-2009] y 3,5\% [2010-2019]. \\

Un alto grado de desigualdad relativa necesariamente lleva a la p\'erdida de la autopercepci\'on de bienestar del grupo de menor ingreso (Obs. (i)), aunque el ingreso sea exponencialmente creciente en el tiempo. En Chile, ver \cite{PNUD2} las cifras de  desigualdad en los ingresos se prueba atrav\'es de algunos indicadores cl\'asicos Gini (G), Indicador de Palma (cociente entre los ingresos del decil m\'as rico y la de 
los cuatro deciles más pobres, D=D10/(D1+D2+D3+D4) y la razón entre el ingreso medio del quintil más rico y más pobre, Q5/Q1. Ver Tabla (\ref{T1}) que muestra sus valores para los a\~nos 1990 y 2013. Respecto a la desigualdad se sabe que aument\'o en [1990, 2000], con una tendencia decreciente en [2000, 2006], y algo m\'as moderada en [2006, 2013].\\
\begin{equation}
\label{T1}
\begin{array}{|c|c|c|c|}
\hline
\mbox{A\~no} & \mbox{G} & \mbox{D} & \mbox{Q5/Q1}  \\
\hline
\hline
1990 & 0,521 & 3,58 & 14,8 \\
2013 & 0,488 & 2,96 &  11,6 \\
\hline
\end{array}    
\end{equation}

Notemos que la hip\'otesis de un indicador de desigualdad relativa $n$ constante, tambi\'en aparece razonable para Chile. En efecto, seg\'un datos extra\'idos de \cite{MK} para  el periodo 1990-2003 el promedio $Q5/Q1=14,5$ con rango $[13,2;15,5]$ y $D10/D1=32,7$ con rango $[27,9; 38,5]$. \\

Como hemos observado, como consecuencia del modelo, alta desigualdad siempre implica descontendo del grupo desfavorecido, podr\'ia ser la situaci\'on de Chile. Claro est\'a,
tasas de crecimiento peque\~nas, hacen crecer el margen de desigualdad relativa en que ambos grupos decaen en su auto percepci\'on.
Adem\'as, solo una tasa de crecimiento suficientemente alta da posibilidad a la existencia de sostener en el tiempo un margen de desigualdad relativa (la que hemos llamado media), que implique el crecimiento del bienestar para ambos grupos.\\ 

En entrevista aparecida en el Diario La Tercera (17/11/2019), el Decano de la Escuela de Artes Liberales, Francisco Covarrubias, acusa un dogmatismo en quienes han administrado el modelo econ\'omico, un error, al ``{\it creer que el óptimo económico siempre coincide con el óptimo social}. Relata que un profesor suyo de la Pontificia Universidad Católica, en sus palabras Chicago Boy emblem\'atico, dec\'ia siempre: ``{\it Lo que importa es que todos tengan un pedazo de la torta y que ese pedazo vaya creciendo. Pero no importa si otro tiene un pedazo más grande}". Nos preguntamos: ¿Qu\'e ocurre cuando se ha instalado socialmente como éxito personal el obtener ``la parte del le\'on"?  \\

Por cierto, el modelo matem\'atico expuesto es b\'asico y no agota el tema y deja la puerta abierta a otras explicaciones. Marco Kremerman, investigador de Fundación Sol, en su estudio: ``{\it Los Bajos Salarios en Chile}", ver \cite{Sol} da cuenta que el 75\% de los trabajadores tiene un ingreso mensual l\'iquido de menos de 
\$500 mil y que si hablamos del 55\% la cifra es menor a \$350 mil. El estudio se\~nala que en Chile ``{\it el costo de vida es muy alto. No se puede tolerar una sociedad en la cual las personas que trabajan asalariadamente a tiempo completo estén en situación de pobreza}". 

\section{Ap\'endice}
\noindent{\bf Demostraci\'on (IV)}:
Si $\lambda^{2}> bb^{*}/(aa^{*})$, entonces $-2aa^{*}\lambda^{2}+4bb^{*}<2aa^{*}\lambda^{2}$. Esto es, $(a-a^{*})^{2}\lambda^{2}+4bb^{*}<(a+a^{*})^{2}\lambda^{2}$. Extrayendo ra\'iz cuadrada y dividiendo a ambos lados por $2b$ tenemos 
$$
\frac{\sqrt{\mathcal{D}}}{2b}<\frac{a\lambda}{2b}+\frac{a^{*}\lambda}{2b}.
$$
Ahora, al sumar $(a-a^{*})\lambda/(2b)$, se obtiene $\hat{n}<a\lambda/b$.\\

Por otro lado, al tener $aa^{*}>bb^{*}/\lambda^{2}$, se obtiene $a/b>b^{*}/(a^{*}\lambda^{2})$. Expresi\'on desde la cual
$$
\frac{b}{b}\frac{b^{*}}{b}> \left( 
\frac{b^{*}}{a^{*}\lambda}\right)^{2} - \frac{b^{*}}{b} \frac{a}{a^{*}}+
\frac{b^{*}}{b}.
$$
Desde lo cual
$$
\frac{(a-a^{*})^{2}\lambda^{2}+4bb^{*}}{4b^{2}}>
 \left( 
\frac{b^{*}}{a^{*}\lambda}\right)^{2} -2\frac{b^{*}}{a^{*}\lambda}\frac{(a-a^{*})\lambda}{2b}+\frac{(a-a^{*})^{2}\lambda^{2}}{4b^{2}}.
$$
A formar un cuadrado de binomio a la derecha para luego extraer ra\'iz cuadrada, se obtiene $\hat{n}>b^{*}/(a^{*}\lambda)$.\\

\noindent{\bf Demostraci\'on (III)}: Se consigue con operaciones muy similares a la prueba de (IV).

\end{document}